\documentclass[manuscript]{aastex}

\shorttitle{[\ion{Fe}{2}] Line Mapping of IC~443}
\shortauthors{Kokusho et al.}

\begin{document}


\title{Large-area [\ion{Fe}{2}] Line Mapping of\\
the Supernova Remnant IC~443 with the IRSF/SIRIUS}


\author{Takuma Kokusho\altaffilmark{1}, Takahiro Nagayama\altaffilmark{1}, Hidehiro Kaneda\altaffilmark{1},\\
Daisuke Ishihara\altaffilmark{1}, Ho-Gyu Lee\altaffilmark{2}, and Takashi Onaka\altaffilmark{2}}

\altaffiltext{1}{Graduate School of Science, Nagoya University, Chikusa-ku, Nagoya 464-8602, Japan}
\altaffiltext{2}{Department of Astronomy, Graduate School of Science, The University of Tokyo, Bunkyo-ku, Tokyo 113-0033, Japan}

\email{kokusho@u.phys.nagoya-u.ac.jp}





\begin{abstract}
We present the result of near-infrared (near-IR) [\ion{Fe}{2}] line mapping of the supernova remnant IC~443 with the IRSF/SIRIUS, using the two narrow-band filters tuned for the [\ion{Fe}{2}] 1.257 $\micron$ and [\ion{Fe}{2}] 1.644 $\micron$ lines.
Covering a large area of 30$\arcmin$ ${\times}$ 35$\arcmin$, our observations reveal that [\ion{Fe}{2}] filamentary structures exist all over the remnant, not only in an ionic shock shell, but also in a molecular shock shell and a central region inside the shells.
With the two [\ion{Fe}{2}] lines, we performed corrections for dust extinction to derive the intrinsic line intensities.
We also obtained the intensities of thermal emission from the warm dust associated with IC~443, using the far- and mid-IR images taken with $AKARI$ and $Spitzer$, respectively.
As a result, we find that the [\ion{Fe}{2}] line emission relative to the dust emission notably enhances in the inner central region.
We discuss causes of the enhanced {[\ion{Fe}{2}]} line emission, estimating the ${\rm Fe}^{+}$ and dust masses.
\end{abstract}






\section{Introduction}

IC~443 is a Galactic supernova remnant (SNR) with an age of 3000${\sbond}$30000 years \citep{ptr88,olb01}.
Existence of a pulsar wind nebula \citep{gns06} and metal-rich X-ray plasma \citep{trj08} indicates that the SNR is of a core-collapse origin.
IC~443 shows a limb-brightened morphology at optical, infrared, and radio wavelengths \citep{byk08,cas11}, while it shows a centrally-peaked morphology in the X-ray \citep{trj08}.
Hence IC 443 is recognized as a mixed-morphology-type SNR.
The distance to IC~443 is estimated to be 1.5 kpc \citep{ws03} and we use this value throughout the present paper.

One of the most interesting and impressive features of IC~443 is its heavy interaction with the interstellar medium (ISM) via ionic and molecular shocks, which makes a marked contrast between the near-infrared (near-IR) morphologies.
The 2MASS survey revealed that IC~443 has a bright shell in the $J$ and $H$ bands in the northeast region, which is faint in the $K_s$ band.
From the shell, the [\ion{Fe}{2}] 1.644 {\micron} line emission was detected \citep{grh87}.
\citet{rho01} concluded that the dominant carriers of both $J$ and $H$ bands are [\ion{Fe}{2}] line emission.
In contrast, the southern ridge is bright in the $K_s$ band but faint in the $J$ and $H$ bands.
The near-IR spectroscopic observation with $AKARI$ detected many ro-vibrational transition lines of ${\rm H}_2$ from parts of the southern ridge \citep{shi11}.
Therefore the ${\rm H}_2$ emission at 2.12 {\micron} is likely to dominate in the $K_s$ band in the southern ridge.
\citet{nor09} showed that the [\ion{Fe}{2}] 26 $\micron$ line emission also exists in the spectra taken along the southern ridge with the $Spitzer$/IRS.
Thus IC~443 is an excellent laboratory to investigate ionic and molecular shock interactions between a SNR and the ambient ISM.

This paper focuses on the spatial distribution of the [\ion{Fe}{2}] line emission in IC~443.
The origin of the [\ion{Fe}{2}] line emission in SNRs is still controversial.
Some mechanism to efficiently produce gas-phase Fe is required because more than 99\% of Fe is usually depleted onto dust grains in the interstellar space \citep{dra95}.
A conventional idea is that the gas-phase Fe is produced by sputtering Fe-bearing dust grains, but we cannot rule out a possibility of supernova ejecta origins.
\citet{koo07} suggested that the [\ion{Fe}{2}] line emission in SNR G11.2-0.3 is partly of an ejecta origin, because, considering its expansion velocity, an ejecta is likely to be excited recently by a reverse shock, where the [\ion{Fe}{2}] line was detected.
To examine the origin of the [\ion{Fe}{2}] line emission, it is important to make a detailed comparison of the spatial distribution of the [\ion{Fe}{2}] line with that of the dust emission associated with IC~443.

\section{Observations and data reduction}

We observed IC~443 with the near-IR camera SIRIUS \citep[Simultaneous Infrared Imager for Unbiased Survey;][]{nas99,nay03} on the IRSF (Infrared Survey Facility) 1.4 m telescope.
IRSF is located at the South African Astronomical Observatory (SAAO), and has been operated by Nagoya University and the SAAO.
The SIRIUS camera has a field of view of 7$\farcm$7 $\times$ 7$\farcm$7 with a pixel scale of 0$\farcs$45.
We observed 44 contiguous fields toward IC~443, which, in total, correspond to the area of 30$\arcmin$ $\times$ 35$\arcmin$, using the two narrow-band filters tuned for the two [\ion{Fe}{2}] lines at 1.257 $\micron$ and 1.644 $\micron$.
The details of the observations are summarized in Table \ref{tbl1}.

We applied the standard data reduction procedure to the array images, including dark subtraction, flat-fielding, sky subtraction, and dithered-image-combining.
As a result, we obtained 44 images for each of the 1.257 $\micron$ and 1.644 $\micron$ filters, which were combined into a large-area map, covering the main structure of IC~443.
We performed photometric calibration, making comparison with the 2MASS point source catalog \citep[PSC; ][]{skr06}.
We used the $J$- and $H$-band magnitudes for the calibrations of the 1.257 $\micron$ and 1.644 $\micron$ filter images, respectively, assuming that the magnitude of each star is the same between the corresponding band and filter images.
For the photometry, we selected stars with fluxes higher than 12.5 mag and flux errors smaller than 0.05 mag from the 2MASS PSC, where we used sufficiently-isolated stars to avoid problems due to star crowding.
The number of the stars used in the photometric calibration is typically 15 per image, and the uncertainties of the photometric calibration coefficients are $\sim$5 \%.
Finally, we subtract point sources with SExtracter \citep{br96} to avoid overestimating the [\ion{Fe}{2}] line intensities due to the stellar flux.

The far-IR images of IC~443 were derived from the $AKARI$ all-sky-survey data.
The $AKARI$ all-sky-survey was performed between May 2006 and August 2007 in the cold mission phase with liquid helium cryogen, using the Far-Infrared Surveyor \citep[FIS;][]{kaw07}.
The FIS has four far-IR photometric bands at central wavelengths of 65 $\micron$ (N60), 90 $\micron$ (WIDE-S), 140 $\micron$ (WIDE-L), and 160 $\micron$ (N160), which have the effective band widths of 22, 38, 52, and 34 $\micron$, respectively.
The data were processed with the $AKARI$ pipeline tool originally optimized for point source extraction \citep{yam09}, and then with an additional pipeline to recover large-scale diffuse emission \citep{doi12}.
We also used the $Spitzer$/MIPS 24 $\micron$ image of IC~443, retrieved from the $Spitzer$ archives (http://archive.spitzer.caltech.edu). 
The observation of IC~443 was part of the LATTER\_SE program (PI: G. Rieke).
The data were taken in a scan map mode \citep{rie04} and produced by the pipeline version S18.12.0.

\section{Results}

Figure \ref{fig1} shows the images of IC~443 with the [\ion{Fe}{2}] 1.257 $\micron$ and 1.644 $\micron$ filters, before the removal of the point sources.
As can be seen in the figure, we detect the [\ion{Fe}{2}] filamentary structures from a wide area of IC~443.
\citet{rho01} detected $J$- and $H$-band diffuse emission with the surface brightnesses of $2.3\ {\times}\ 10^{-4}$ and $1.6\ {\times}\ 10^{-4}\ {\rm ergs}\ {\rm s}^{-1}\ {\rm cm}^{-2}\ {\rm sr}^{-1}$, respectively, from a 48$\arcsec$ $\times$ 18$\arcsec$ area in the northeast shell, centered at (R.A., Decl.) = ($06^{\rm h}17^{\rm m}34{\fs}41$, $+22{\degr}52{\arcmin}55{\farcs}2$).
At the same position, \citet{grh87} detected the [\ion{Fe}{2}] 1.644 $\micron$ line intensity of $1.1\ {\times}\ 10^{-4}\ {\rm ergs}\ {\rm s}^{-1}\ {\rm cm}^{-2}\ {\rm sr}^{-1}$ , and our maps show the [\ion{Fe}{2}] 1.257 $\micron$ and 1.644 $\micron$ line intensities of $1.7\ {\times}\ 10^{-4}$ and $1.2\ {\times}\ 10^{-4}\ {\rm ergs}\ {\rm s}^{-1}\ {\rm cm}^{-2}\ {\rm sr}^{-1}$, respectively.
Thus in the northeast shell, indeed, the [\ion{Fe}{2}] 1.257 $\micron$ and the 1.644 $\micron$ line emission are dominant in the $J$ and $H$ band, respectively.
More importantly, the [\ion{Fe}{2}] filaments are detected not only in the northeast shell, which is bright in the $J$ and $H$ bands, but also in other large area including the southern ridge and a central region inside the shells.
For the entire remnant, \citet{rho01} obtained the $J$- and $H$-band fluxes of  $1.2\ {\times}\ 10^{-9}$ and $6.4\ {\times}\ 10^{-10}\ {\rm ergs}\ {\rm s}^{-1}\ {\rm cm}^{-2}$, respectively, while we detected the [\ion{Fe}{2}] 1.257 $\micron$ and 1.644 $\micron$ line fluxes of  $8.2\ {\times}\ 10^{-10}$ and $6.2\ {\times}\ 10^{-10}\ {\rm ergs}\ {\rm s}^{-1}\ {\rm cm}^{-2}$, respectively.

To obtain the intrinsic [\ion{Fe}{2}] line intensities, we performed corrections for foreground dust extinction, using the observed line ratio of the two [\ion{Fe}{2}] lines at 1.257 $\micron$ and 1.644 $\micron$.
Since they are due to electronic transitions from the same upper level, the intrinsic line ratio must be fixed at 1.36 \citep{ns88}.
We estimated the foreground extinction by comparing the observed line ratio with the theoretical one, assuming the interstellar extinction law of \citet{car89}.
In calculating the line ratios, we first masked the pixels where [\ion{Fe}{2}] $1.257\ \micron$ and $1.644\ \micron$ are below 1${\sigma}$, because the fluctuation of background can produce strong positive or negative spikes in a ratio map.
Then we smoothed the resultant ratio map with a boxcar of 150 $\times$ 150 pixels.
The extinction map thus derived is shown in the left panel of Fig. \ref{fig2} on the color scale, with the contours of the $^{12}{\rm CO}$ map integrated in the velocity range of $-10$ to 0 km ${\rm s}^{-1}$ from \citet{lee12}.
From the figure, it can be seen that the foreground extinction is large in the west and southeast regions, and CO clouds are lying in similar regions where \citet{trj06} showed the absorption of X-rays. 
Using the extinction map, we performed the extinction corrections for both 1.257 $\micron$ and 1.644 $\micron$ images.
The resultant [\ion{Fe}{2}] 1.257 $\micron$ line intensity image is shown in the right panel of Fig. \ref{fig2} on the gray scale.

The far- and mid-IR images of IC~443, which are obtained by $AKARI$ and $Spitzer$, respectively, include emission not only from the dust associated with IC~443, but also from foreground and background dust.
Therefore discrimination of these two components is required.
The dust associated with IC~443 is heated by shocks and therefore likely to have higher temperatures than foreground and background dust.
We do not consider the dust component associated with IC~443, but not yet heated by shocks, if any, because it cannot be the origin of the observed ${\rm Fe}^{+}$ through dust sputtering.  
Hence we fitted mid- to far-IR spectral energy distributions (SEDs) by a two-temperature modified blackbody model with the emissivity power-law index of unity.
First, we fitted the model to the SED of the total region with a photometry radius of 19$\arcmin$, centered at (R.A., Decl.) = ($06^{\rm h}17^{\rm m}18{\fs}63$, $+22{\degr}37{\arcmin}07{\farcs}5$).
As a result, we find that the model with the temperatures of $T_{\rm cold}=14.8$ K and $T_{\rm warm}=56.3$ K reproduces the total SED very well.
Then with the temperatures fixed at the above best-fit values, we fitted local SEDs for every spatial bin of $15{\arcsec}\ {\times}\ 15{\arcsec}$, allowing only the amplitudes of the two components to vary.
Here, considering possible severe contaminations of the [\ion{O}{1}] 63 $\micron$ line emission to the 65 $\micron$ narrow-band (N60) intensity and [\ion{C}{2}] 158 $\micron$ line emission to the 160 $\micron$ narrow-band (N160) intensity, we excluded the N60 and N160 band intensities in the local SED fitting.
The contours in the right panel of Fig. \ref{fig2} show the resultant distribution of the intensity of the warm dust emission integrated over the wavelength range of 2$\sbond$200 $\micron$, which is derived from the above SED fitting.

Here and hereafter we define regions A to D, as shown in the right panel of Fig. \ref{fig2}, all of which have the same area of 2.1 ${\times}$ 2.1 ${\rm pc}^2$ (4${\farcm}$9 ${\times}$ 4${\farcm}$9).
Regions A, centered at (R.A., Decl.) = ($06^{\rm h}17^{\rm m}55{\fs}91,\ +22{\degr}46{\arcmin}24{\farcs}8$) and D at (R.A., Decl.) = ($06^{\rm h}17^{\rm m}40{\fs}58,\ +22{\degr}23{\arcmin}55{\farcs}0$) represent the ionic shock and the molecular shock region, respectively, while regions B at (R.A., Decl.) = ($06^{\rm h}17^{\rm m}13{\fs}62,\ +22{\degr}41{\arcmin}10{\farcs}0$) and C at (R.A., Decl.) = ($06^{\rm h}17^{\rm m}21{\fs}75,\ +22{\degr}32{\arcmin}10{\farcs}0$) represent inner central regions where the [\ion{Fe}{2}] line emission is detected significantly.
As can be seen in the figure, both [\ion{Fe}{2}] and warm dust emission are strong in region A, while only the warm dust emission is strong in region D.
In regions B and C, the [\ion{Fe}{2}] line emission is relatively strong whereas the dust emission is faint.
Near region C, \citet{nor09} also detected the [\ion{Fe}{2}] 26 $\micron$ emission in their $Spitzer$/IRS spectrum.
More quantitatively, Fig. \ref{fig3} shows the correlation plot between the intensities of the extinction-corrected [\ion{Fe}{2}] 1.257 $\micron$ line and the warm dust emission integrated over the wavelength range of 2$\sbond$200 $\micron$.
For comparative purpose, we draw the line fitted to only the data points of region A, which gives the slope of 0.058 $\pm$ 0.006 and the correlation coefficient of 0.3.
The figure clearly reveals that the [\ion{Fe}{2}] line emission in regions B and C is notably strong as compared with the warm dust emission.

\section{Discussion}
We estimate the masses of ${\rm Fe}^{+}$ and warm dust for regions A to D.
In a two-level approximation, the [\ion{Fe}{2}] 1.644 $\micron$ line intensity is given by
\begin{equation}
I_{\lambda1.644}=\frac{T_4^{-0.94}e^{-1.57/T_4}(N_{\rm Fe^{+}}/10^{16})}{1+4.2\ {\times}\ 10^4T_4^{0.69}/n_e},
\end{equation}
where $T_4$ is the gas temperature in units of $10^4$ K, and $n_e$ and $N_{\rm Fe^{+}}$ are the electron density and the ${\rm Fe}^{+}$ column density in units of ${\rm cm}^{-3}$ and ${\rm cm}^{-2}$, respectively \citep{bli94}.
Assuming $T_4=1.2$ and $n_e=500$, which were derived from the mid-IR [\ion{Fe}{2}] lines \citep{rho01}, we estimate $N_{\rm Fe^{+}}$ and thus the ${\rm Fe}^{+}$ mass, $M_{{\rm Fe}^{+}}$, from the area of each region.
As a result, $M_{{\rm Fe}^{+}}$ is 5, 1, 0.9, and 0.8 ${\times}\ 10^{-4}\ M_{\sun}$ for regions A to D.
Note that, if we adopt the typical values derived from the near-IR [\ion{Fe}{2}] lines in SNRs, $T_4=0.5$ and $n_e=1\ {\times}\ 10^4$ \citep[e.g.,][]{grh87, koo07}, $M_{{\rm Fe}^{+}}$ decreases by an order of magnitude, while if we adopt the gas density of only a few times 10 ${\rm cm}^{-3}$ which was suggested by \ion{H}{1} \citep{lee08} and optical line observations \citep{Fes80}, $M_{{\rm Fe}^{+}}$ increases by an order of magnitude.
Accordingly the above ${\rm Fe}^{+}$ masses can have such systematic errors.

On the other hand, by assuming the dust absorption coefficient given by \citet{hld83}, a grain radius of 0.1 $\micron$, and a specific dust mass density of 3 g ${\rm cm}^{-3}$, the mass of the warm dust, $M_{\rm d}$, is expressed as
\begin{equation}
M_{\rm d}=10^4\left(\frac{L_{\rm warm}}{10^8L_{\sun}}\right) {\left(\frac{T_{\rm warm}}{40\ {\rm K}}\right)}^{-5}\  M_{\sun},
\end{equation}
where $T_{\rm warm}$ and $L_{\rm warm}$ are the temperature and luminosity of the warm dust, respectively, derived from the above SED fitting.
As a result, we obtain $M_{\rm d}$ of 50, 2, 0.9, and 50 ${\times}\ 10^{-4}\ M_{\sun}$ for regions A to D.
Hence the $M_{{\rm Fe}^{+}}/M_{\rm d}$ ratios are 0.1, 0.5, 1, and 0.02 for regions A to D.
To first order, the absorption coefficient of a silicate grain is proportional to its volume, and thus the mass estimation is not affected by the grain size, although the shock processing of dust will change the grain size distribution \citep[e.g.,][]{and11}.
{If stochastic heating of very small grains is dominant, the dust masses could be overestimated.

The above huge differences in $M_{{\rm Fe}^{+}}/M_{\rm d}$ can be interpreted by the differences in the ratio of the destroyed to the non-destroyed dust mass from region to region.
We estimate the latter ratios to be 30, 70, 80, and 8 \% for regions A to D, assuming that all elements in dust are removed away by sputtering at the same rate, all gas-phase Fe is singly ionized, and Fe initially depleted to dust accounts for 22 \% of $M_{\rm d}$ \citep{dra07}.
Accordingly, the initial dust masses are 70, 7, 5 and 50 ${\times}\ 10^{-4}\ M_{\sun}$ for regions A to D.
The difference between regions A and D can be explained by the fact that more destructive shock is propagating in region A than in region D where the deceleration of the shock due to the molecular cloud is indicated \citep{rho01}.
Our result shows that the spatial distribution of the initial dust grains is highly biased toward regions A and D, probably reflecting the distribution of the pre-existing ISM.

In regions B and C, the dust seems to be almost completely destroyed, which might be reasonable considering that the shock front reached the interior regions earlier.
The sputtering destruction time of dust in the hot plasma of IC~443 is estimated to be ${\sim}1\ {\times}\ 10^5$ yr \citep{ds79}, by using the plasma density and temperature of 1 ${\rm cm}^{-3}$ and $10^7$ K \citep{ptr88}, respectively, and assuming the initial grain radius of 0.1 $\micron$.
This time scale is, however, considerably longer than the age of IC~443.
\citet{jon94} showed that dust destruction associated with a J-shock is much more efficient, which is suitable for regions A and D, and could also be viable for regions B and C.
On the other hand, the time for Fe to reach an ionization equilibrium in hot plasma, ${\sim}10^{12}$ $n_e^{-1}$ s, where $n_e$ is the electron density of X-ray plasma \citep{mas94}, is estimated to be ${\sim}2\ {\times}\ 10^4$ yr, by using $n_e$ of 1.7 ${\rm cm}^{-3}$ obtained for IC~443 by \citet{yag09}.
Since this time scale is comparable to the age of IC~443, a significant fraction of gas-phase Fe atoms is likely to be more highly ionized than ${\rm Fe}^{+}$ in regions B and C, whereby we may underestimate the ${\rm Fe}^{+}$ masses (i.e., the destroyed dust masses) there.

The ${\rm Fe}^{+}$ mass in the central region of IC~443 is estimated to be $2\ {\times}\ 10^{-3}\ M_{\sun}$ from the total [\ion{Fe}{2}] line flux in the observed area relatively free of the pre-existing ISM, where the intensity of the warm dust emission is lower than $1.5\ {\times}\ 10^{-3}\ {\rm ergs}\ {\rm s}^{-1}\ {\rm cm}^{-2}\ {\rm sr}^{-1}$ (i.e., the lowest contour level in the right panel of Fig. \ref{fig2}). 
According to the model calculation, the $^{56}{\rm Fe}$ mass produced from a 15$\sbond$25 $M_{\sun}$ core-collapse supernova is 0.05$\sbond$0.13 $M_{\sun}$ \citep{thi96}, which is one to two orders of magnitude larger than the ${\rm Fe}^{+}$ mass derived from our estimation.
Thus our result suggests that Fe ejecta may not be so abundant as predicted theoretically for a core-collapse supernova.

\section{Conclusion}
With the IRSF/SIRIUS, we observed the Galactic SNR IC~443, using the narrow-band filters for the [\ion{Fe}{2}] 1.257 $\micron$ and 1.644 $\micron$ lines.
Our observations detected the [\ion{Fe}{2}] line emission from a wide area of IC~443, not only the shock-heated shell regions, but also interior regions, with prominent filamentary, shell-like structures.
From a combination of the [\ion{Fe}{2}] 1.257 $\micron$ and 1.644 $\micron$ lines which are attributed to electronic transitions from the same upper energy level, we estimated the foreground extinction and obtained extinction-corrected [\ion{Fe}{2}] line maps.
We also obtained the map of the shock-heated warm dust emission using $AKARI$ and $Spitzer$ IR images.
As a result, we find that the [\ion{Fe}{2}] line emission is notably strong relative to the warm dust emission in the inner central part of IC~443.
We estimated the ${\rm Fe}^{+}$ masses from the intensity of [\ion{Fe}{2}] line and compared them with the masses of the shock-heated warm dust, which are derived from the SED fitting by using the $AKARI$ and $Spitzer$ IR images.
The result implies that the mass ratios of ${\rm Fe}^{+}$ to the dust are 0.1 (ionic shock shell), 0.02 (molecular shock shell), and 0.5 and 1 (interior regions).
The large difference in the ratio likely reflects that in the dust destruction efficiency; in particular, the ${\rm Fe}^{+}$/dust enhancement in the interior regions requires an efficient destruction mechanism for large grains, or contribution of theoretically-predicted Fe ejecta.

We thank the anonymous referee for giving us helpful comments to improve the paper.
The IRSF project was financially supported by the Sumitomo foundation and Grants-in-Aid for Scientific Research on Priority Areas (A) (Nos. 10147207 and 10147214) from the Ministry of Education, Culture, Sports, Science and Technology (MEXT).
The operation of IRSF is supported by Joint Development Research of National Astronomical Observatory of Japan, and  Optical Near-Infrared Astronomy Inter-University Cooperation Program, funded by the MEXT of Japan.
This research is based on observations made with $AKARI$, a JAXA project with the participation of ESA, and with $Spitzer$, which is operated by the Jet Propulsion Laboratory, California Institute of Technology under a contract with NASA.

\begin{deluxetable}{cccccc}
\tablecaption{Summary of the present IRSF observations \label{tbl1}}
\tablewidth{0pt}
\tablehead{
\colhead{${\lambda}_{\rm center}$ ($\micron$)} & \colhead{Transition} & \colhead{${\Delta}{\lambda}_{\rm eff}$\tablenotemark{a} ($\micron$)} &
\colhead{Exposure (s)} & \colhead{NDI\tablenotemark{b}} & \colhead{Date}
}
\startdata
1.257 & $a^{4}D_{7/2}\ {\rightarrow}\ a^{6}D_{9/2}$  & 0.028 & 60 & 10 & Nov 2012\\
1.644 & $a^{4}D_{7/2}\ {\rightarrow}\ a^{4}F_{9/2}$  & 0.018 & 30 & 15 & Feb, Nov 2012\\ 
\enddata
\tablenotetext{a}{Effective band width calculated from ${\int}S({\lambda})d{\lambda}=T_{\lambda}{\Delta}{\lambda}_{\rm eff}$, where S($\lambda$) is a filter response curve, and $T_{\lambda}$ is a throughput at ${\lambda}_{\rm center}$.}
\tablenotetext{b}{Number of dithered images.}
\end{deluxetable}

\begin{figure}
\plotone{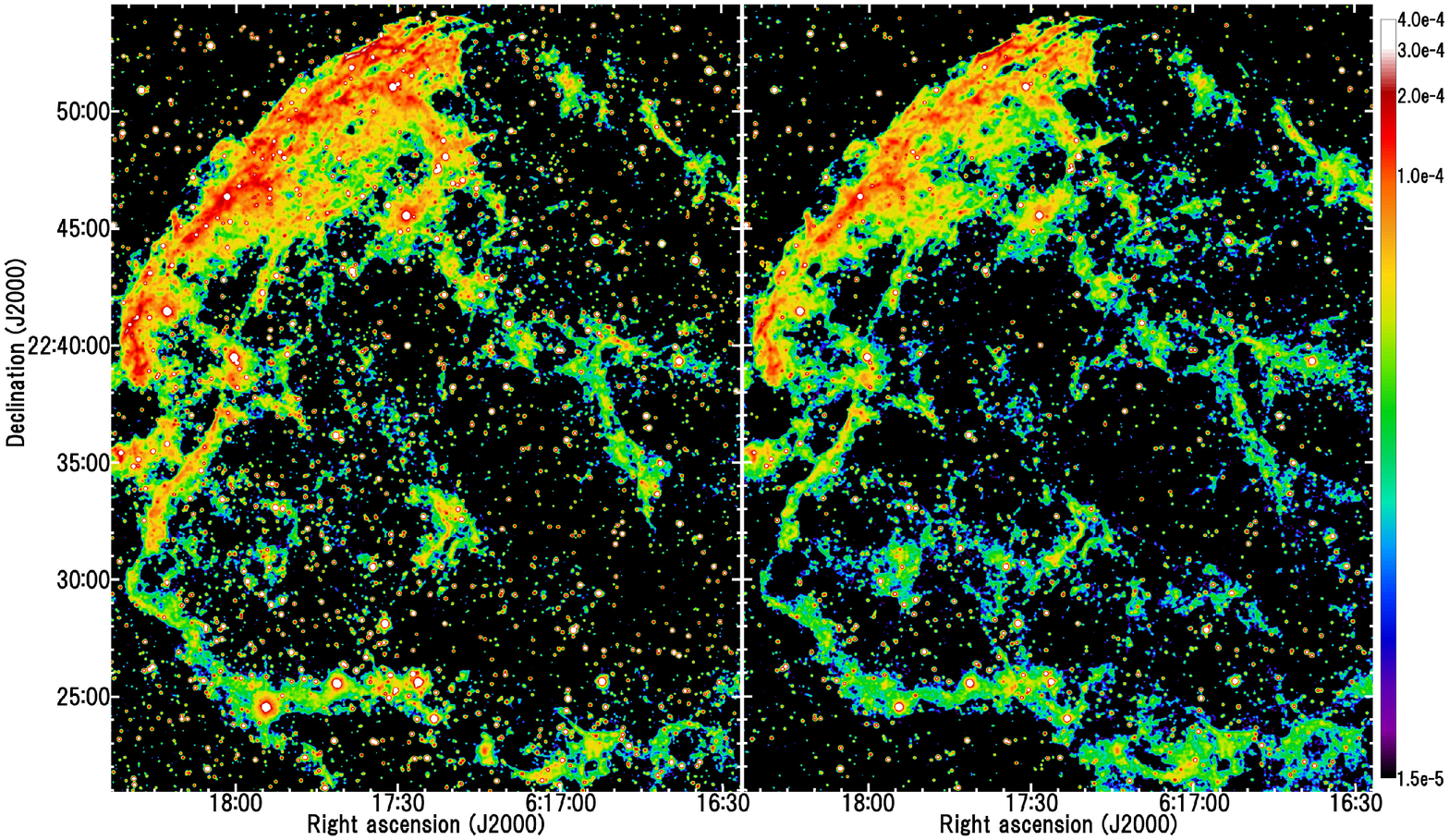}
\caption{
[\ion{Fe}{2}] 1.257 $\micron$ (left) and 1.644 $\micron$ (right) line maps, smoothed with a Gaussian kernel of 2.3$\arcsec$ in sigma.
The color levels are given in units of ${\rm ergs}\ {\rm s}^{-1}\ {\rm cm}^{-2}\ {\rm sr}^{-1}$.
\label{fig1}}
\end{figure}

\begin{figure}
\plotone{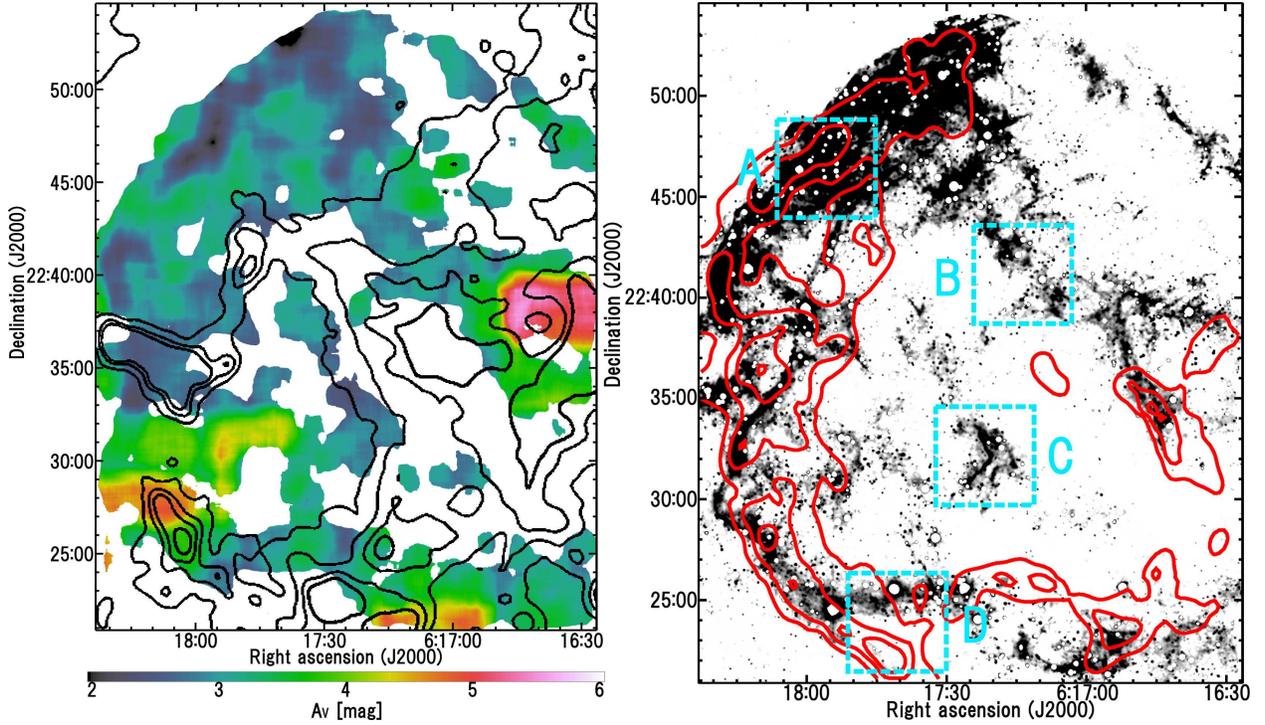}
\caption{
(Left) Foreground dust extinction $A_V$ map derived from the observed [\ion{Fe}{2}] line intensities, with the contours of the $^{12}{\rm CO}$ map integrated in the velocity range of $-10$ to 0 km ${\rm s}^{-1}$ taken from \citet{lee12}.
(Right) Spatial distribution of the extinction-corrected [\ion{Fe}{2}] 1.257 $\micron$ line intensity, with the contours of the intensity of the warm dust emission integrated over the wavelength range of 2$\sbond$200 $\micron$, which is derived from the SED fitting.
The gray scale ranges from 2.0 to 8.0 ${\times}\ 10^{-5}\ {\rm ergs}\ {\rm s}^{-1}\ {\rm cm}^{-2}\ {\rm sr}^{-1}$, while the contour levels are 1.5, 2.5, and 3.5 ${\times}\ 10^{-3}\ {\rm ergs}\ {\rm s}^{-1}\ {\rm cm}^{-2}\ {\rm sr}^{-1}$.
\label{fig2}}
\end{figure}

\begin{figure}
\plotone{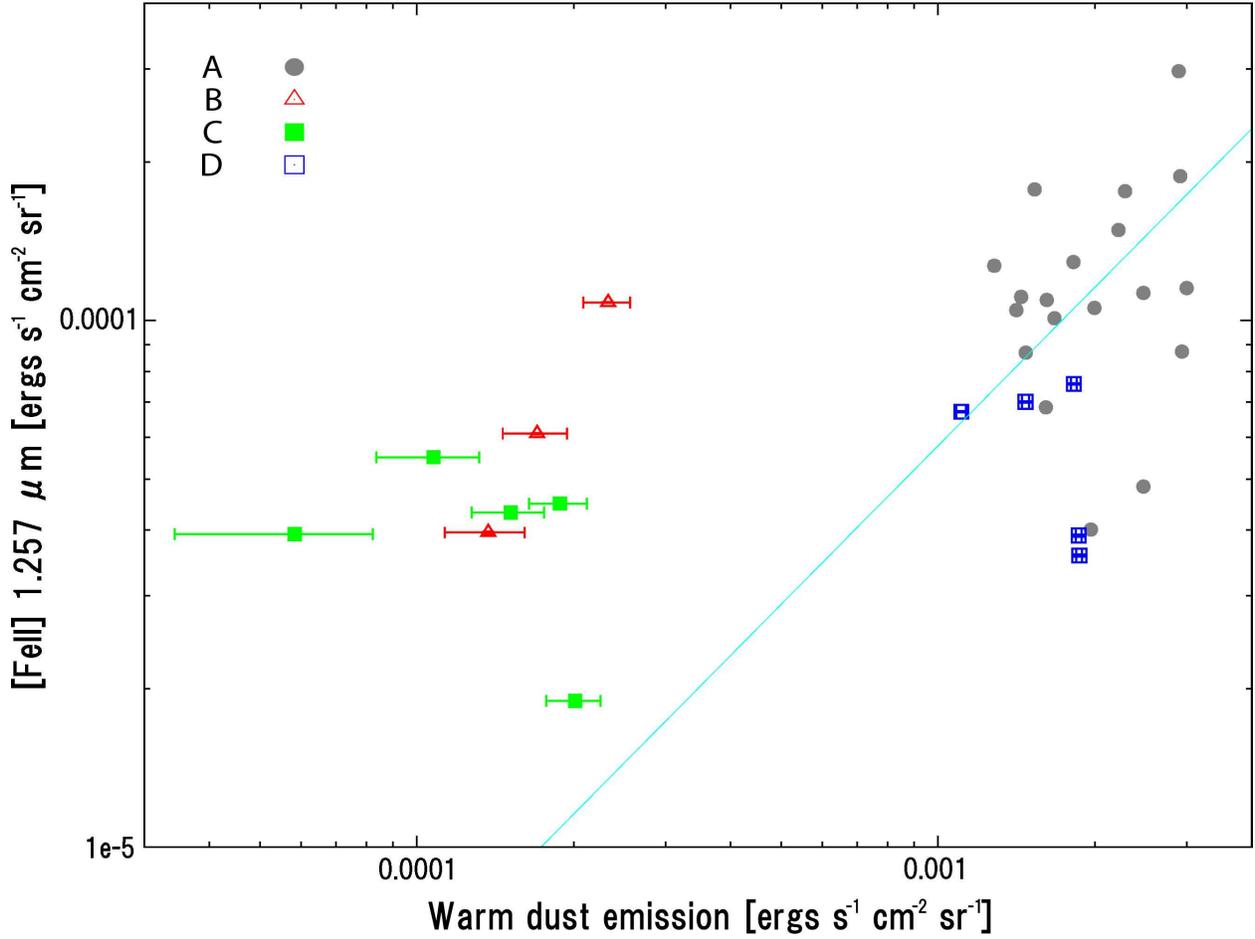}
\caption{
Correlation plot between the intensities of the extinction-corrected [\ion{Fe}{2}] 1.257 $\micron$ line and the warm dust emission integrated over the wavelength range of 2$\sbond$200 $\micron$.
Regions A to D are defined in Fig. \ref{fig2}, and the line fitted to only the data points of region A is overlaid.
The data are spatially sampled every 1$\arcmin$, which corresponds to 130 pixels of the camera array.
\label{fig3}}
\end{figure}


\begin{thebibliography}{}
\bibitem[Andersen et al.(2011)]{and11} Andersen, M., Rho, J., Reach, W.~T., Hewitt, J.~W., \& Bernard, J.~P.\ 2011, \apj, 742, 7
\bibitem[Bertin \& Arnouts(1996)]{br96} Bertin, E., \& Arnouts, S.\ 1996, \aaps, 117, 393
\bibitem[Blietz et al.(1994)]{bli94} Blietz, M., Cameron, M., Drapatz, S., et al.\ 1994, \apj, 421, 92
\bibitem[Bykov et al.(2008)]{byk08} Bykov, A.~M., Krassilchtchikov, A.~M., Uvarov, Y.~A., et al.\ 2008, \apj, 676, 1050
\bibitem[Cardelli et al.(1989)]{car89} Cardelli, J.~A., Clayton, G.~C., \& Mathis, J.~S.\ 1989, \apj, 345, 245
\bibitem[Castelletti et al.(2011)]{cas11} Castelletti, G., Dubner, G., Clarke, T., \& Kassim, N.~E.\ 2011, \aap, 534, A21
\bibitem[Doi et al.(2012)]{doi12} Doi, Y., Komugi, S., Kawada, M., et al.\ 2012, PKAS, 27, 111
\bibitem[Draine \& Salpeter(1979)]{ds79} Draine, B.~T., \& Salpeter, E.~E.\ 1979, \apj, 231, 77
\bibitem[Draine(1995)]{dra95} Draine, B.~T.\ 1995, \apss, 233, 111
\bibitem[Draine et al.(2007)]{dra07} Draine, B.~T., Dale, D.~A., Bendo, G., et al.\ 2007, \apj, 663, 866
\bibitem[Fesen \& Kirshner(1980)]{Fes80} Fesen, R.~A., \& Kirshner, R.~P.\ 1980, \apj, 242, 1023
\bibitem[Gaensler et al.(2006)]{gns06} Gaensler, B.~M., Chatterjee, S., Slane, P.~O., et al.\ 2006, \apj, 648, 1037
\bibitem[Graham et al.(1987)]{grh87} Graham, J.~R., Wright, G.~S., \& Longmore, A.~J.\ 1987, \apj, 313, 847
\bibitem[Hildebrand(1983)]{hld83} Hildebrand, R.~H.\ 1983, \qjras, 24, 267
\bibitem[Jones et al.(1994)]{jon94} Jones, A.~P., Tielens, A.~G.~G.~M., Hollenbach, D.~J., \& McKee, C.~F.\ 1994, \apj, 433, 797
\bibitem[Kawada et al.(2007)]{kaw07} Kawada, M., Baba, H., Barthel, P.~D., et al.\ 2007, \pasj, 59, 389
\bibitem[Koo et al.(2007)]{koo07} Koo, B.-C., Moon, D.-S., Lee, H.-G., Lee, J.-J., \& Matthews, K.\ 2007, \apj, 657, 308
\bibitem[Lee et al.(2008)]{lee08} Lee, J.-J., Koo, B.-C., Yun, M.~S., et al.\ 2008, \aj, 135, 796
\bibitem[Lee et al.(2012)]{lee12} Lee, J.-J., Koo, B.-C., Snell, R.~L., et al.\ 2012, \apj, 749, 34
\bibitem[Masai(1994)]{mas94} Masai, K.\ 1994, \apj, 437, 770
\bibitem[Nagashima et al.(1999)]{nas99} Nagashima, C., Nagayama, T., Nakajima, Y., et al. 1999, in Star Formation 1999, ed. T. Nakamoto (Nagano: Nobeyama Radio Obs.), 397
\bibitem[Nagayama et al.(2003)]{nay03} Nagayama, T., Nagashima, C., Nakajima, Y., et al.\ 2003, \procspie, 4841, 459
\bibitem[Noriega-Crespo et al.(2009)]{nor09} Noriega-Crespo, A., Hines, D.~C., Gordon, K., et al.\ 2009, The Evolving ISM in the Milky Way and Nearby Galaxies
\bibitem[Nussbaumer \& Storey(1988)]{ns88} Nussbaumer, H., \& Storey, P.~J.\ 1988, \aap, 193, 327
\bibitem[Olbert et al.(2001)]{olb01} Olbert, C.~M., Clearfield, C.~R., Williams, N.~E., Keohane, J.~W., \& Frail, D.~A.\ 2001, \apjl, 554, L205
\bibitem[Petre et al.(1988)]{ptr88} Petre, R., Szymkowiak, A.~E., Seward, F.~D., \& Willingale, R.\ 1988, \apj, 335, 215
\bibitem[Rieke et al.(2004)]{rie04} Rieke, G.~H., Young, E.~T., Engelbracht, C.~W., et al.\ 2004, \apjs, 154, 25
\bibitem[Rho et al.(2001)]{rho01} Rho, J., Jarrett, T.~H., Cutri, R.~M., \& Reach, W.~T.\ 2001, \apj, 547, 885
\bibitem[Shinn et al.(2011)]{shi11} Shinn, J.-H., Koo, B.-C., Seon, K.-I., \& Lee, H.-G.\ 2011, \apj, 732, 124
\bibitem[Skrutskie et al.(2006)]{skr06} Skrutskie, M.~F., Cutri, R.~M., Stiening, R., et al.\ 2006, \aj, 131, 1163
\bibitem[Thielemann et al.(1996)]{thi96} Thielemann, F.-K., Nomoto, K., \& Hashimoto, M.-A.\ 1996, \apj, 460, 408
\bibitem[Troja et al.(2006)]{trj06} Troja, E., Bocchino, F., \& Reale, F.\ 2006, \apj, 649, 258
\bibitem[Troja et al.(2008)]{trj08} Troja, E., Bocchino, F., Miceli, M., \& Reale, F.\ 2008, \aap, 485, 777
\bibitem[Welsh \& Sallmen(2003)]{ws03} Welsh, B.~Y., \& Sallmen, S.\ 2003, \aap, 408, 545
\bibitem[Yamaguchi et al.(2009)]{yag09} Yamaguchi, H., Ozawa, M., Koyama, K., et al.\ 2009, \apjl, 705, L6
\bibitem[Yamamura et al.(2009)]{yam09} Yamamura, I., Makiuti, S., Ikeda, N., et al.\ 2009, in ASP Conf. Ser. 418, AKARI, a Light to Illuminate the Misty Universe, ed. T. Onaka et al. (San Francisco, CA: ASP), 3
\end{thebibliography}
\end{document}